\documentclass[review]{elsarticle}

\usepackage{lineno,hyperref}
\usepackage{multirow}
\usepackage{multicol}
\usepackage{tabularx}
\usepackage{makecell}
\usepackage{verbatim}
\usepackage{array}
\usepackage{longtable}

\usepackage{soul}
\usepackage[margin=1in,letterpaper]{geometry} %
\usepackage[table,x11names]{xcolor}
    
\modulolinenumbers[5]

\journal{Journal of \LaTeX\ Templates}

\begin{document}

\begin{frontmatter}

\title{Feature Extraction Using Deep Generative Models for Bangla Text Classification on a New Comprehensive Dataset}


\author[address1]{Md. Rafi-Ur-Rashid \corref{mycorrespondingauthor}}
\cortext[mycorrespondingauthor]{Corresponding author}
\ead{mur5028@psu.edu}
\author[address2]{Sami Azam}
\author[address2]{Mirjam Jonkman}

\address[address1]{Department of Computer Science and Engineering, Penn State University, State College, PA 16802, United States}
\address[address2]{College of Engineering, IT, and Environment, Charles Darwin University, Casuarina, NT 0810, Australia}

\begin{abstract}
Selection of features for text classification is a fundamental task in text mining and information retrieval. Despite being the sixth most widely spoken language in the world, Bangla has received little attention due to the scarcity of text datasets. In this research, we collected, annotated, and prepared a comprehensive dataset of \textbf{212,184} Bangla documents in seven different categories and made it publicly accessible. We implemented three deep learning generative models: LSTM variational autoencoder (LSTM VAE), auxiliary classifier generative adversarial network (AC-GAN), and adversarial autoencoder (AAE) to extract text features, although their applications are initially found in the field of computer vision. We utilized our dataset to train these three models and used the feature space obtained in the document classification task. We evaluated the performance of the classifiers and found that the adversarial autoencoder model produced the best feature space.
\end{abstract}


\end{frontmatter}

\section{Introduction}
Textual feature extraction refers to the task of finding latent properties from a text. The aim is to select a set of features by employing an effective way to reduce the dimensions of the feature space. It also removes noisy, uncorrelated, or superfluous features from the feature space, thereby improving the accuracy and efficiency of the learning algorithm. In practice, many feature extraction methods require carefully handcrafted features through a lengthy and erroneous process. However, deep learning algorithms do not have these issues as the key aspect of deep learning is that the features are not designed by human engineers but learned from data using a learning procedure. Deep learning requires minimal engineering by hand, so it can easily take advantage of an increase in the number of available computations and data. However, working with deep learning algorithms has some inherent challenges. The availability of a large volume of data is an essential prerequisite to train deep learning models. Although several large datasets on other languages are available, the existing text corpus on Bangla contains only a small collection of documents. The primary challenge is that to ensure the availability of quality data. Many of the available Bangla texts on the web contain several grammatical, semantic, and spelling mistakes. Another challenge is that the amount of labeled text data is very scarce. In most cases, texts are unlabeled, and they are a mix of data from different categories. Hence in practice, the collected data need to be labeled manually with the help of data annotators. Due to these challenges, so far only a few studies have utilized deep learning models for NLP tasks in Bangla. To deal with the limitations described above, we have developed a large comprehensive text corpus of \textbf{212,184} Bangla articles which we utilized to train, validate, and test the feature extraction models. Five human annotators manually labeled the articles in seven categories: Government \& Politics, Science \& Technology, Economics, Health \& Lifestyle, Entertainment, Arts \& Literature, and Sports. We made this dataset publicly accessible for future research on Bangla language processing.\\

Here we experiment with three deep learning feature extraction models: LSTM variational autoencoder (LSTM VAE), auxiliary classifier generative adversarial network (AC-GAN), and adversarial autoencoder (AAE). The variational autoencoder (VAE) is a generative model that can be considered a normal autoencoder combined with variational inference. It encodes data to latent (random) variables and then decodes them to reconstruct the data. Generative Adversarial Network, or GAN, is an architecture for training generative models. It is frequently used in computer vision. However, this class of algorithms is still under-utilized in the field of NLP. In this work, we attempt to transform the AC-GAN model \citep{odena2017conditional} for analyzing textual data rather than pixel values of an image, and we use it for dimensionality reduction of a vector space. The adversarial autoencoder is a network inspired by the ideas behind variational autoencoders, merging them into GAN concepts. Like the GANs, its general applications are in computer vision. Here we use it for working with text. 

Since the generation capability of these deep learning generative models has been already proven in computer vision, we wanted to observe their performance and applicability for natural language processing. Hence, we got motivated to experiment with these models in this work. The feature space obtained from these models is used in the document classification task. We then make a comparative analysis of the performance of these three feature extraction algorithms and discuss the pros and cons of each algorithm. We also applied Principal component analysis (PCA) and Bidirectional Encoder Representations from Transformer (BERT) \cite{devlin2018bert} to analyze their performance against our proposed models. The major contributions of this work are summarized below:
\begin{itemize}
  \item We prepare a comprehensive human-annotated dataset of \textbf{212,184} Bangla articles from seven different categories. To the best of our knowledge, it is the largest collection of documents with the highest number of categories in Bangla language so far. We also made this corpus available for public access.
  \item We implement three deep learning generative models: LSTM VAE, AC-GAN, and adversarial autoencoder to extract textual features.
  \item We use our curated dataset to train, validate, and test the three feature extraction models and utilize the feature space obtained for the document classification task.
  \item Finally, we made a comparative analysis on the performance of the models mentioned above along with a couple of state-of-the-art methods.
\end{itemize}

\section{Related Work}
\label{rel}
Feature extraction plays a crucial role in text classification. Existing text feature extraction methods include Word embedding techniques like Word2Vec~\citep{ling2015two}, Glove~\citep{pennington2014glove}, term frequency-based methods such as Tf-idf ~\citep{yun2005improved}, mapping methods like latent semantic analysis \citep{dumais2004latent}, principal component analysis \citep{abdulhussain2015experimental} and several clustering methods such as concept indexing \citep{kim2005dimension}, and CHI \citep{uysal2012novel}.

Deep learning models emerged in the field of research and industry from early 2000, primarily due to the substantial advantages these learning models offer, such as automatic feature engineering, the ability to work with unstructured data, high-quality results with acceptable efficiency, and so on. At present, deep learning feature representation includes autoencoder, restricted Boltzmann model, deep belief network, convolutional neural network \citep{lopez2017deep} and recurrent neural network \citep{elman1990finding}, etc. Contextualized word embeddings such as BERT \cite{devlin2018bert} is also being used for text representation and summarization tasks. For example, \cite{li2019automatic} utilized BERT embeddings in combination with BiLSTM and CNN output features for automatic text classification. They targeted to improve the accuracy of text classification task by transforming the text-to-dynamic character-level embedding using BERT and making full use of CNN to extract the local features as well as using BiLSTM to have the advantage of memory to link the extracted context features. Although their proposed method greatly improved in accuracy, the time cost problem of the model did not effectively improve. Besides, a BERT features based algorithm is used by \citep{xu2020bert} for predicting the helpfulness scores of online customers reviews. In this study, they developed a neural network (NN) based model with BERT features, instead of explanatory variables to rank the Amazon product reviews dataset. Their proposed BERT based NN model overcame time consuming process associated with selecting and extracting explanatory variables and derived better results in terms of mean absolute error and standard deviation. However, due to limited computing power, the test could not be run using the best hyper parameters.

Restricted Boltzman Machine (RBM), a type of machine learning tool with strong power of representation, has been utilized as the feature extractor in a large variety of classification problems \citep{cai2012feature}. For instance, \citep{liu2010novel} proposed a novel text classification approach based on deep belief network (DBN), a deep neural network constituted by multiple layers of RBM. Moreover, \citep{sun2014chinese} presents a DBN model and a multi-modality feature extraction method to extend features from short text for Chinese microblogging sentiment classification. Although, with proper structure and parameter, the performance of DBN in text classification is better than the state-of-the-art surface learning models (e.g., SVM or NB), it is not as good performing as the other sophisticated DL architectures like CNNs and RNNs. 

CNN is one of the artificial neural networks, with its strong adaptability and good performance at mining local characteristics of data. Therefore, \citep{kalchbrenner2014convolutional} adopted a Dynamic Convolutional Neural Network (DCNN) for the semantic modelling of sentences. Their proposed network handles input sentences of varying length and induces a feature graph over the sentence that can explicitly captures word relations of varying size. Besides, a similar experiment was done by \citep{ma2015dependency} that proposed a simple dependency based convolution framework to exploit the non-local interactions between words and their method outperformed sequential CNN baselines on modeling sentences. Apart from that, \citep{rakhlin2016convolutional} sketched several typical CNN models applied to feature extraction in text classification, and filtered with different lengths, which were used to convolve text matrix. In convention, RNNs and its improved variants like LSTMs \citep{hochreiter1997long} are used to process sequential data. In \citep{lai2015recurrent} LSTM unites with CNN to gain vectorization representation of the whole sentence. For minimizing noise of traditional window-based neural network when learning word representations, they applied a recurrent structure to capture contextual information as far as possible. Their model outperformed CNN and
Recursive NN with respect to four different text classification datasets.

Another trend in deep learning is the autoencoder model, first introduced by  \cite{rumelhart1985learning}. It is a feedforward network that can learn a compressed, distributed representation of data, usually with the goal of dimensionality reduction. It has a wide range of applications, including semantic parsing, handwritten digit recognition \citep{shopon2016bangla} and feature extraction \citep{liu2014feature}. Besides, deep autoencoders have been also used for processing network security and medical data \citep{sparse}. VAE is a variant of the autoencoder model. Standard autoencoders are not ideal for the task of generation. They convert the input data into the encoded vector, which lies in a latent space. On the other hand, the variational autoencoder not only reconstructs the input but also has the ability to generate data similar to the input. VAE has been frequently applied in NLP tasks. For example, \citep{semeniuta2017hybrid} implemented a novel VAE model for text generation. Unlike other existing works, where both encoder and decoder layers are LSTMs, the core of their model is a feed-forward architecture composed of one-dimensional convolutional and deconvolutional layers. They empirically verified that their model is easier to train and more effective in converging on longer texts than its fully recurrent alternatives. Apart from that,\citep{kruengkrai2019better} came up with a new idea of sampling latent variables multiple times at a gradient step which helps in improving a variational autoencoder. Later, they proposed a simple and effective method that can serve as a strong baseline for latent variable text modeling. Furthermore, a deep generative model of bilingual sentence pairs is proposed by \citep{eikema2020map} for machine translation. They employed a framework of amortized variational inference and derived an efficient approximation to MAP decoding that requires only a
single forward pass through the network for prediction. GANs are another form of generative models which are frequently applied in the image domain \citep{pan2019recent}. More recently, however, there has been an increasing amount of research dedicated to their text applications. For instance, \citep{haidar2019latent} introduced a novel text-based approach called Soft-GAN as a new solution for the main bottleneck of using GAN for text generation. Later, their proposed techniques showed superiority over other conventional GAN-based
techniques. Besides, \citep{rajeswar2017adversarial} addressed the discrete output
space problem of GAN by simply forcing the discriminator to operate on continuous valued output distributions. They also showed that it is possible to perform conditional generation of text on high-level sentence features such as sentiment and questions. Apart from that, the adversarial autoencoder \citep{makhzani2015adversarial} architecture combines the variational autoencoders with the concept of GAN. It has gathered a lot of attention from the computer vision community, yielding impressive results for image generation. \citep{zhang2017age} proposed a conditional adversarial autoencoder (CAAE) that learns a face manifold for realizing smooth age progression and regression simultaneously. Although, these deep generative models yield appealing results with the image data, they are still under-utilized in the natural language processing domain. Hence, we thought of implementing LSTM VAE, AC-GAN, and adversarial autoencoder for extracting text features.

The application of deep learning algorithms in NLP research in Bangla has been limited. \cite{hassan2016sentiment} utilized a deep learning recurrent model, specifically, Long Short Term Memory for sentiment analysis, on Bangla and Romanized Bangla text. NER (Named Entity Recognition) using Word2Vec and BERT embeddings for Bangla Language was performed by \cite{ashrafi2020banner}. Apart from that, \cite{alam2016bidirectional} designed a system using LSTM neural networks followed by Conditional Random Fields (CRFs) for Bangla POS tagging. The problem of Bangla sentence correction and auto-completion was addressed by  \cite{islam2018bangla}. They used a decoder-encoder-based sequence-to-sequence recurrent neural network. To the best of our knowledge, no work to analyze feature extraction from Bangla texts using deep learning generative models has been done so far. In this work, we implement three deep learning feature extraction models and subsequently apply them to Bangla document classification.

\section{Materials and Methods}
\label{exp}
The overall methodology of our work is depicted in  Figure \ref{overview}.

\subsection{Dataset}
\label{dataset_desc}
To collect a comprehensive amount of texts, we select some of the most reputed \footnote{https://www.alexa.com \\  We used the Alexa ranking to determine the popularity of a website.} online Bangla news portals and blogs. In addition, we also collected texts from several comparatively less popular sources in order to increase variation in the dataset. We used a customized python web crawler to scrape those articles. While collecting articles from these sources, we found that many sites have the same content. We, therefore, discarded duplicates and irrelevant texts. Besides, some of the documents were very lengthy, containing a large number of words. We split those documents into multiple smaller instances and thus further increased the number of instances in our corpus. We then labeled each article in one of seven categories: Government \& Politics, Science \& Technology, Economics, Health \& Lifestyle, Entertainment, Arts \& Literature, and Sports. Five trained human annotators performed the labeling task. They labeled each article individually by analyzing its content, title, source, and other metadata. Finally, we counted the majority vote for the annotations to assign a label to each document. The distribution of these articles into seven categories is shown in Table \ref{corpus details}\footnote{Copyright \& Fair Use Acknowledgement: The contents used here fall under the "Fair Use Policy" which allows ``fair use'' for purposes such as criticism, comment, news reporting, teaching, scholarship, and research.}. Full description of the dataset with sources can be found in \ref{corpusA} and \ref{sourceA}.
\begin{table}[!t]
\centering
\small{
\begin{tabular}{c|c|c|c|c}

     \hline
       \multirow{2}{*}{Category}& No. of &Avg. \# characters &Avg. \# words &Avg. \# sentences\\
       & documents & per document & per document & per document\\
      \hline
      Govt. \& Politics & 62,974 & 1927 & 138 & 11\\
    
      Science \& Technology & 11,004 & 3108& 180 & 14\\
     
      Economics & 20,922 & 1764&129 & 9\\
      
      Health \& Lifestyle & 13,467 & 3443&184 & 13\\
      
      Entertainment & 34,702 & 2924&175 & 11\\
      
      Arts \& Literature & 15,214 & 5917&339 & 27\\
      
      Sports & 53,901 & 3423&186 & 14\\
      \hline
      
\end{tabular}
\caption{Corpus details}
 \label{corpus details}
 }
\end{table}

\subsection{Text Preprocessing}
Before extracting features from the text, some preprocessing steps are necessary. First, punctuation marks, digits, and special characters are removed. Next, each document is tokenized using blank spaces that results in a list of words. We have developed a list of \textbf{430} Bangla stopwords that have no relevance in analyzing the documents, rather than increasing ambiguity during the learning session of our models. Hence we excluded those stop words from the texts. Finally, we did stemming on some of the frequently used Bangla words \citep{stemm} to reduce the complexity of the feature space.

The preprocessed words (i.e., unstructured data) require numeric representations that will be fed into the Deep learning models as input data. We utilize word embeddings ~\citep{journals/corr/abs-1301-3781} for this purpose. Word embeddings provide a dense (vector) representation of words and their relative meanings. We adopt an Embedding layer approach where the embeddings are learned jointly as part of our deep learning model. For this purpose, each preprocessed word is one-hot encoded and then fed to the embedding layer. The output of this layer is a vector of a certain length for each word. We specified the embedding dimension for each document and the number of words for the embedding vectors to 128 and 200, respectively.
\begin{figure}[h]
\includegraphics[width=\linewidth]{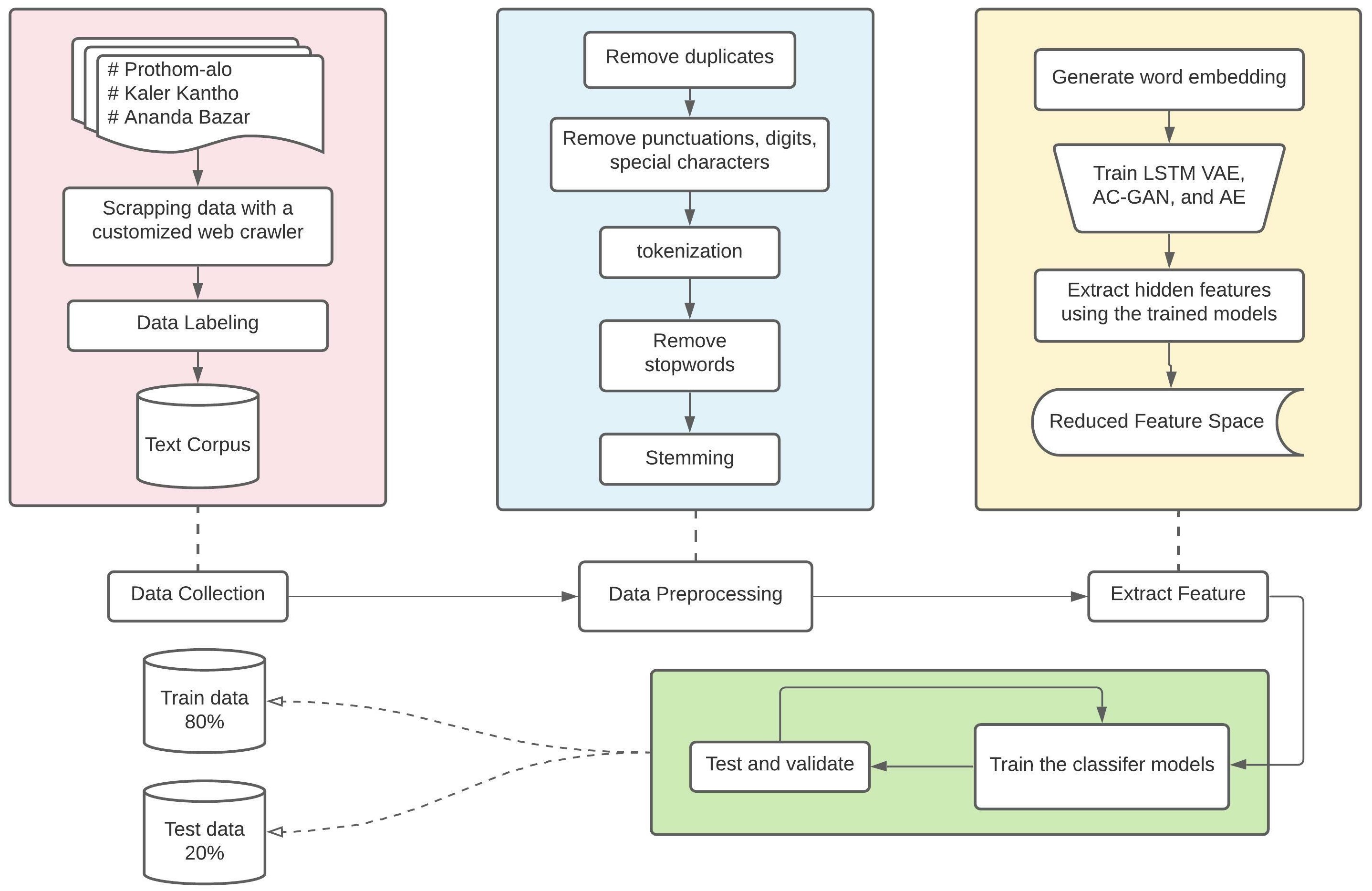} 
\caption{Overview of the Methodology}
\label{overview}
\end{figure}
\subsection{Dataset Splitting}
 The dataset is split into a 70\% training set, a 10\% validation set, and a 20\% testing set. We ensure that adequate data variation prevails in all three subsets and maintain uniformity among the classes. The train, validation, and testing sets are populated in such a way that the proportion of texts from the seven categories is the same for each of them.
\begin{figure}[h]
\centering
\includegraphics[width=\linewidth]{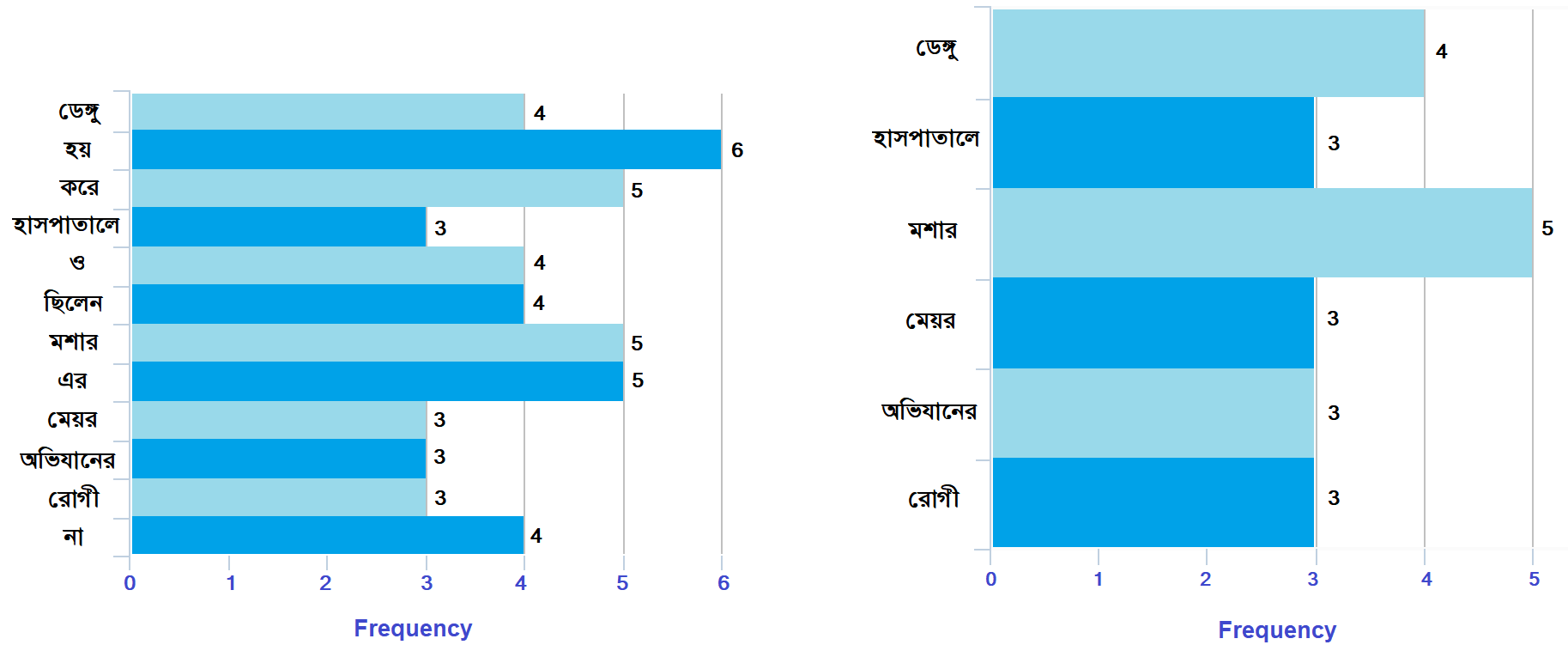} 
\caption{Sample list of words before and after stopword removal}
\label{overview}
\end{figure}
\subsection{Feature Extraction Models}
After the preprocessing step, each document is represented as a $[200, 128]$ dimensional vector. This feature space is too large and may cause overfitting and poor accuracy during the text classification task. We, therefore, aim to extract a more compact feature space that will be fed into the document classifier model later. We implemented three deep learning generative models for this purpose: LSTM VAE, AC-GAN, and adversarial autoencoder. An overview of each of those models will be given in the following subsubsections.

\subsubsection{LSTM Variational Autoencoder}
In regular deterministic autoencoders, the latent code does not learn the probability distribution of the data, making it unsuitable for replicating new data. The VAE solves this problem by explicitly defining a probability distribution in the latent code. In fact, it learns the latent representations of the inputs not as single points but as soft ellipsoidal regions in the latent space. To achieve it, the model is trained by maximizing the variational lower bound on the data log-likelihood under the generative model. 

\begin{figure}[!t]
\centering
\includegraphics[width=\linewidth]{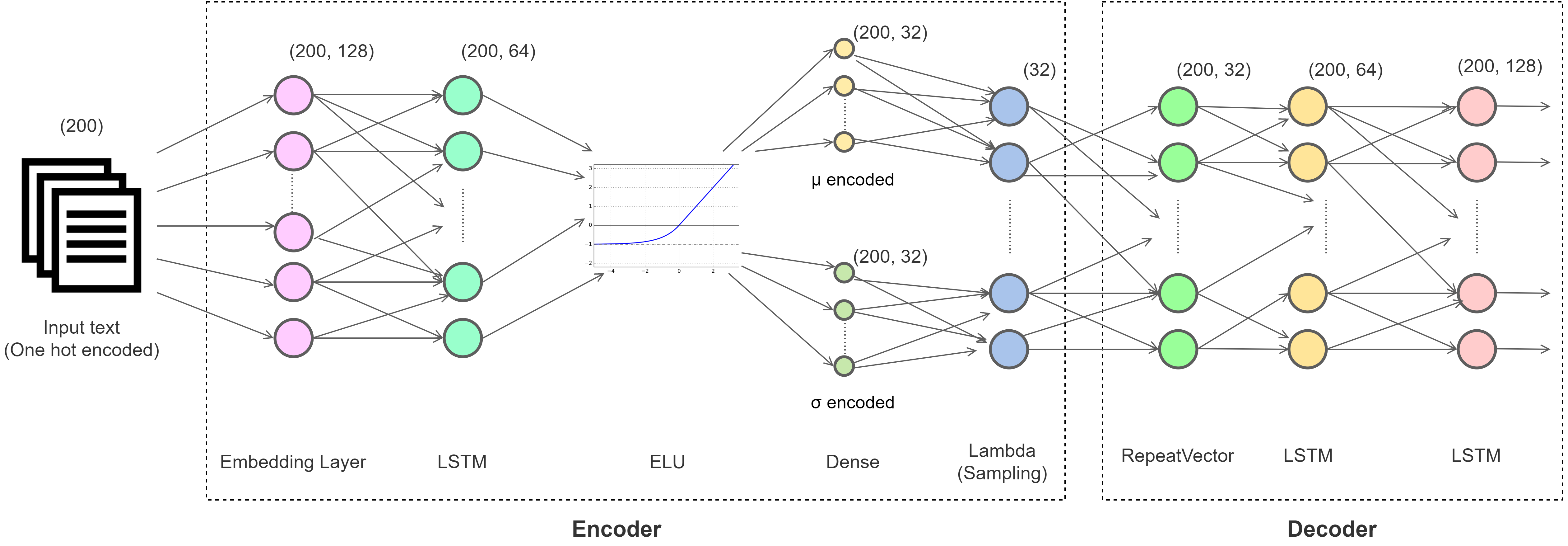} 
\caption{The LSTM VAE model}
\label{vae}
\end{figure}

 We tried to replicate the VAE model proposed by \cite{bowman2015generating}. It is composed of one recurrent LSTM encoder network, a fully connected network for the variational inference, and two Recurrent LSTM decoder networks. The entire network is trained simultaneously via SGD. Figure \ref{vae} shows the detailed architecture. After the input is passed through the LSTM and ELU activation layers, it is encoded into two parts, the mean value - $ \mu$ and standard deviation - $ \sigma$. These are sampled to create a sampled encoding vector which is passed to the decoder. In the decoder part, this encoded input is repeated multiple times and then passed through the two LSTM layers, which decode it into the original dimension of the input embedding, i.e., 128. In this way, VAE creates output that points to the area where the encoded value can be. The mean value controls the point where the center of the encoding is located, and the standard deviation defines the area where the encoding can vary from the mean. This approach gives us encoded vectors which are as close to each other as possible. To summarize the VAE model:\\
\textbf{Encoder:}
\begin{itemize}
\item Input: Embedding.
\item Output: Sampled encoding vector.    
\end{itemize}
\textbf{Decoder:}
\begin{itemize}
    \item Input: Sampled encoding vector.
    \item Output: Reconstructed embedding.
\end{itemize}

We trained the model for 100 epochs, with a batch size of 256. We used the Adam optimizer with an initial learning rate of 0.001 and applied the custom variational loss function proposed in the original work. This basically minimizes the \textbf{Kullback–Leibler divergence} (KL divergence). We then extracted the features from the sampled encoder output, which has the compressed embedding dimensions of $[200,32]$ .

\subsubsection{AC-GAN}
\begin{figure}[!t]
\centering
\includegraphics[scale=0.85]{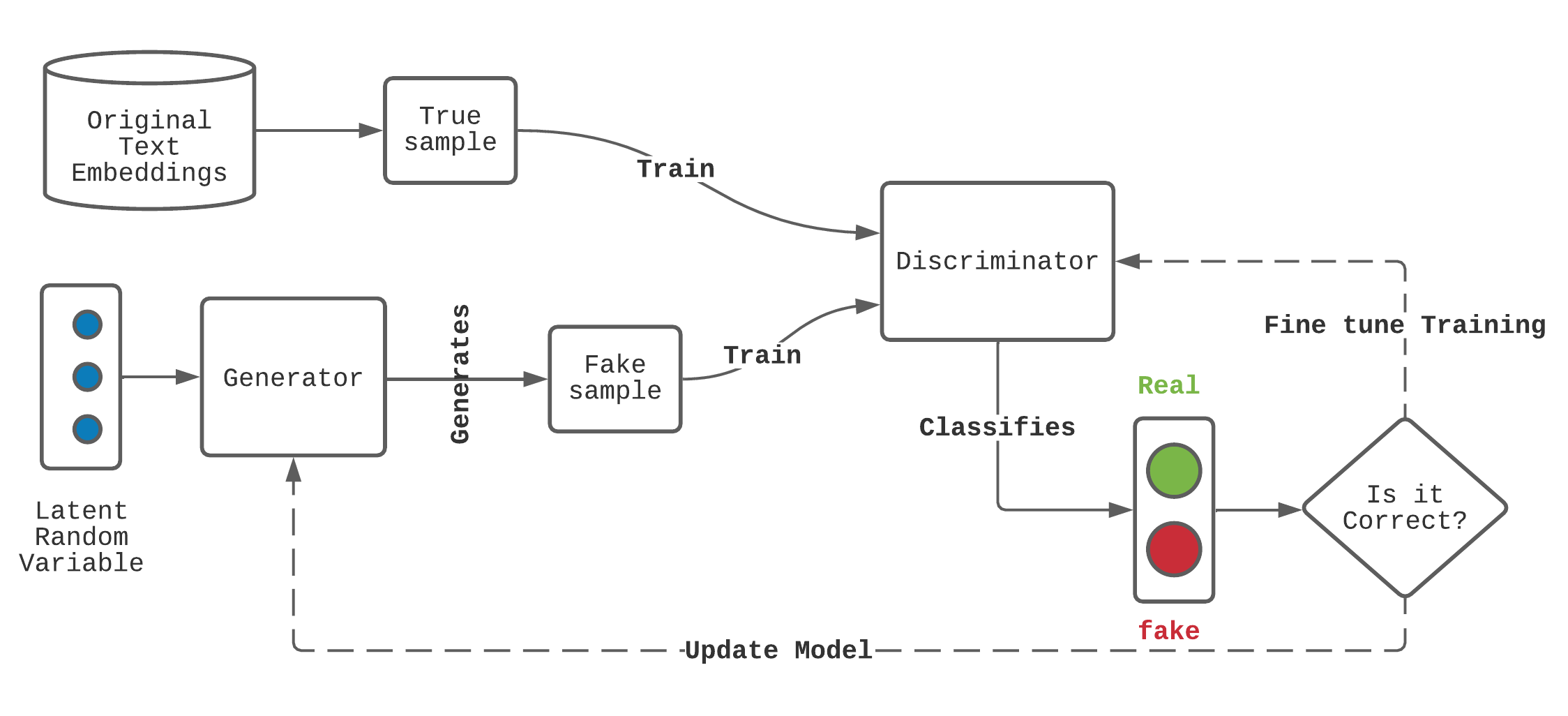} 
\caption{Workflow of a Basic GAN for embedding synthesis}
\label{gan}
\end{figure}

The conditional generative adversarial network is a type of GAN that involves the conditional generation of target vectors. Training of the GAN model is modified in such a way that the generator is provided both with a point in the latent space and a class label as input, generating a vector for that class. The discriminator is provided with both a vector and a class label and must classify whether the vector is real or fake. The Auxiliary Classifier GAN, or AC-GAN for short, is an extension of the conditional GAN that changes the discriminator to predict the class label of a given vector rather than receiving it as input. It has the effect of stabilizing the training process whilst learning a representation in latent space that is independent of the class label. Figure \ref{gan} illustrates the workflow of a basic GAN model for synthesizing embeddings. To summarize: \\
\textbf{Generator model:}
\begin{itemize}
\item Input: Random point from latent space, and class label.
\item Output: Generated embedding.    
\end{itemize}
\textbf{Discriminator model:}
\begin{itemize}
    \item Input: Embedding.
    \item Output: Probability that the provided embedding is real, probability of the embedding belonging to each known class.
\end{itemize}

\begin{figure}[!t]
\includegraphics[width=\linewidth]{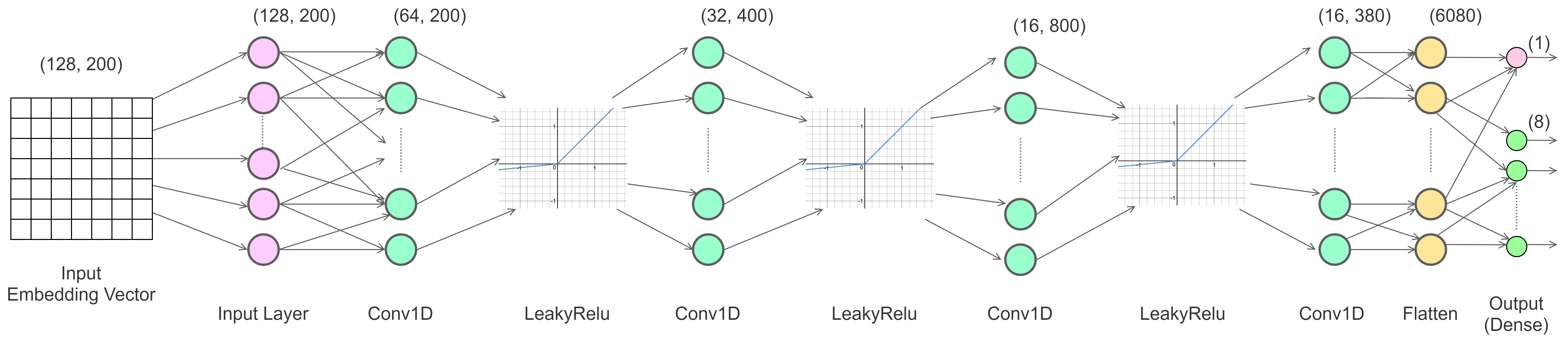} 
\caption{The discriminator model of AC-GAN}
\label{disc}
\end{figure}

The discriminator model of the AC-GAN that we implemented in this work can be described as the DC-GAN architecture \citep{radford2015unsupervised} with a few modifications. These include using Gaussian weight initialization, batch normalization, LeakyReLU, Dropout, and a stride of size 2 for downsampling instead of pooling layers. Since we are processing sequences (text embeddings) instead of grid data (image), we utilized 1D convolution layers, not 2D. Figure  \ref{disc} provides a simplified illustration (Dropout and Batch normalization layers are not shown) of the discriminator model. It has two output layers. The first output layer consists of a single node with the sigmoid activation for predicting the realness of the embedding. The second output layer consists of multiple nodes, one for each class, using the softmax activation function to predict the class label of the given embedding. The model is trained with two loss functions: binary cross-entropy for the first output layer and categorical cross-entropy loss for the second output layer. The Adam optimization is used with a learning rate of 0.002 to fit the model.

\begin{figure}[!t]
\centering
\includegraphics[width=\linewidth]{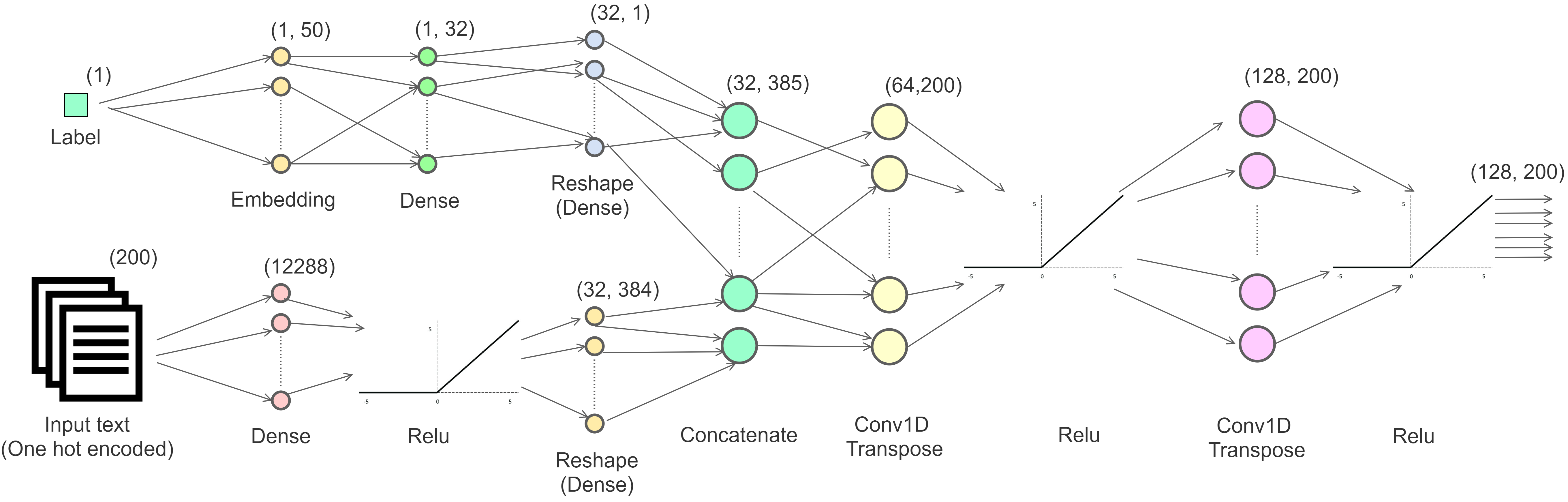} 
\caption{The generator model of AC-GAN}
\label{gen}
\end{figure}

The architecture of the generator model is shown in Figure \ref{gen}. It takes a random point from the latent space and the class label as input, then outputs a generated embedding of dimension $[128,200]$. The point in latent space is interpreted by a fully connected layer with sufficient activations to create a vector dimension which is a multiple of 32, in this case, 384. Rather than the one-hot encoding of the seven class labels, we represent them as a learned embedding with an arbitrary number of dimensions, 50 in this case. The output of this embedding layer is then interpreted by a fully connected layer with a linear activation function. This is then concatenated with the latent input, resulting in 385 feature maps. These feature maps are passed through two transpose convolutional layers to upsample the 32-dimensional feature maps, first to 64 and then finally to 128 dimensions. Again, we implemented 1D transpose convolution instead of 2D for processing the sequences. The 1D transpose convolutional layers are basically formed by stacking one Lambda layer to expand the tensor dimension, one 2D transpose convolutional layer for upsampling, and another Lambda layer, to squeeze the tensor dimension to the same level as before.

We built a composite model that stacks the generator model on top of the discriminator model. It ensures that the generator model is not updated directly; instead, it is updated via the discriminator model. The input to this composite model is the input to the generator model, namely a random point from the latent space and a class label. The generator model is connected directly to the discriminator model, taking the generated embedding directly as input. The discriminator model predicts both the realness of the generated embedding and the class label. We performed the training for 5000 epochs with a batch size of 200. After the training, features from the second convolutional layer output of the discriminator model, an embedding of dimension $[32,200]$, were extracted and reshaped into $[200,32]$.

\subsubsection{Adversarial Autoencoder}
\begin{figure}[!t]
\centering
\includegraphics[width=\linewidth]{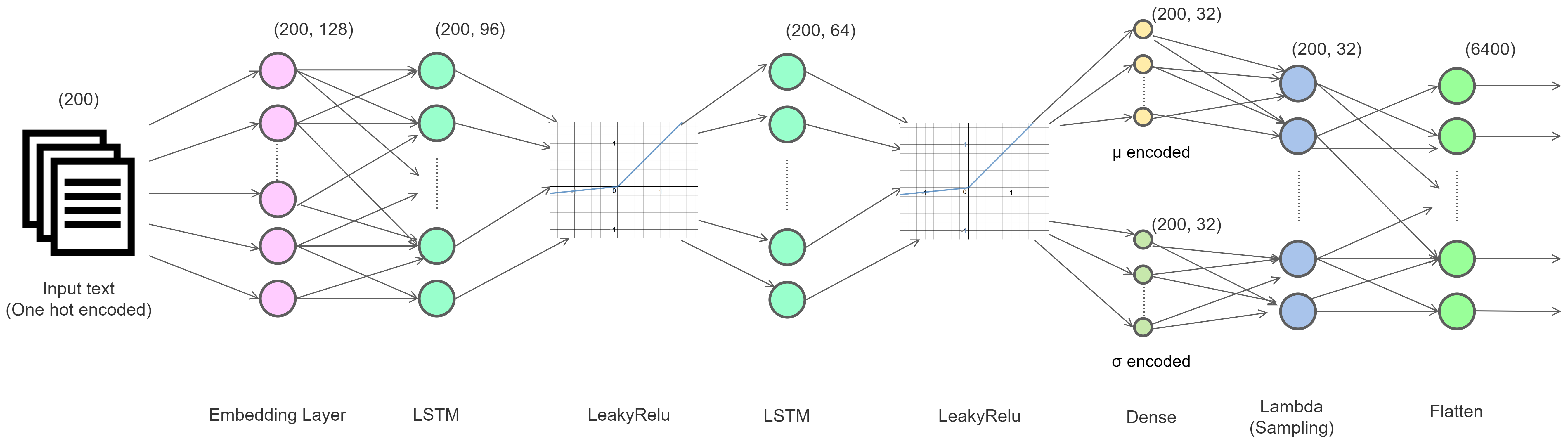} 
\caption{The Encoder model of Adversarial Autoencoder}
\label{aae_encoder}
\end{figure}

\begin{figure}[!t]
\includegraphics[width=\linewidth]{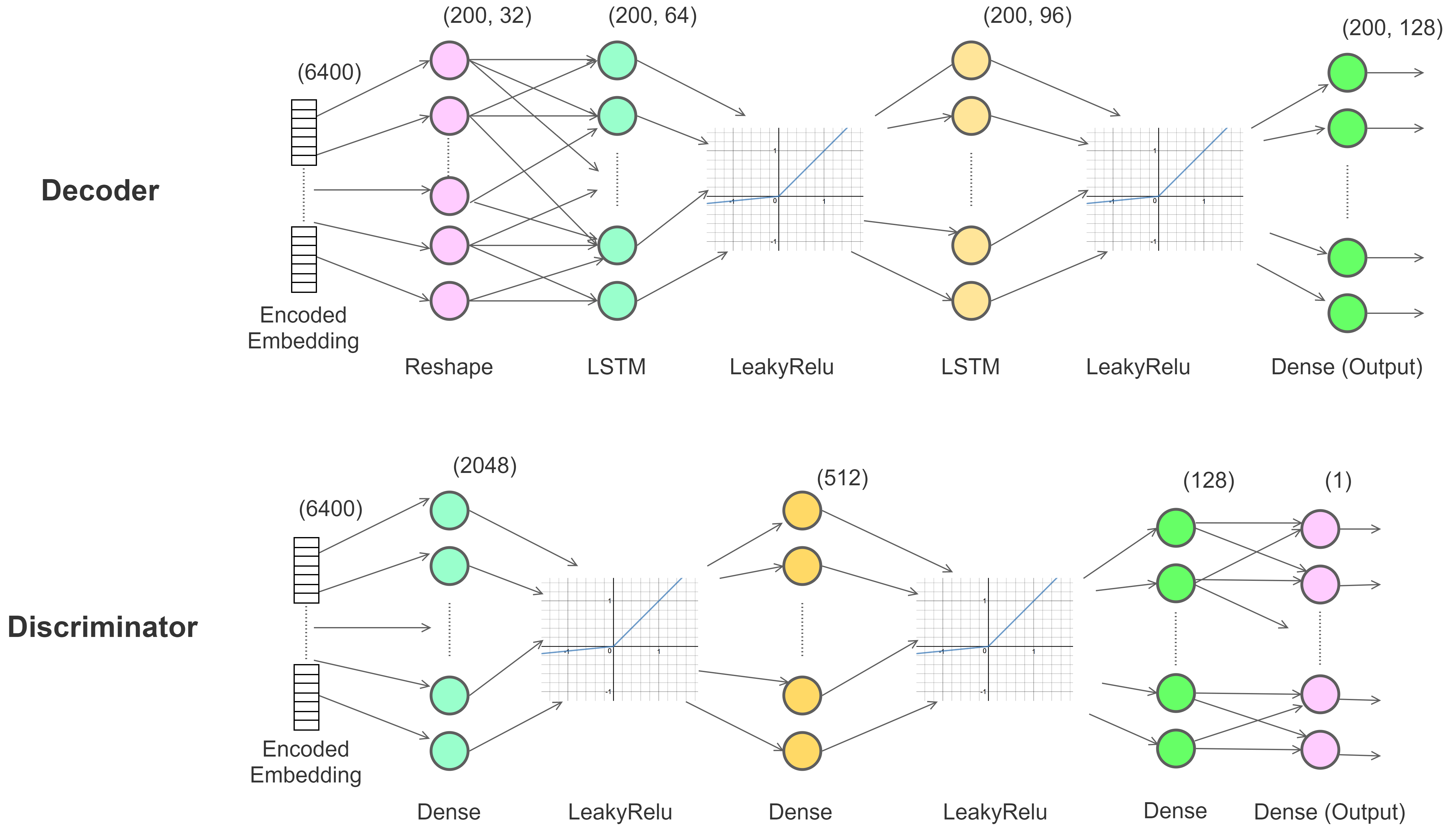} 
\caption{The Decoder \& the discriminator model of Adversarial Autoencoder}
\label{aae_dec_disc}
\end{figure}

Adversarial Autoencoder (AAE) has a similar purpose for continuous encoded data as VAE. The encoded vector is again composed of the mean value and the standard deviation, but it now uses a \textbf{prior distribution} to control the encoder output. Instead of minimizing the KL-divergence between latent codes distribution and the desired distribution, it uses a discriminator to discriminate latent codes and samples from the desired distribution. Any adversarial autoencoder model has three distinct parts: the encoder, the decoder, and the discriminator. In this work, the encoder model was implemented with two LSTM layers. The remainder of the encoder is identical to the VAE model, except for the activation function, which in this case is the Leaky rectifier.  Figure \ref{aae_encoder} shows the encoder architecture. For the decoder, we used the same architecture as for the encoder. The sigmoid function was used as the activation for the output layer. The function of the discriminator is to take the encoded representation and distinguish fake embeddings from the real ones so that the output is one neuron. Activations used are LeakyRelu for the two hidden layers and sigmoid for the output layer. The detailed architecture of the decoder and the discriminator model are shown in Figure \ref{aae_encoder} and \ref{aae_dec_disc}.
To summarize:\\
\textbf{Encoder(Generator):}
\begin{itemize}
\item Input: Embedding.
\item Output: Latent representation of the embedding, i.e. sampled encoding vector with latent dimension.    
\end{itemize}
\textbf{Decoder:}
\begin{itemize}
    \item Input: Encoded representation of the embedding.
    \item Output: Reconstructed embedding.
\end{itemize}
\textbf{Discriminator:}
\begin{itemize}
    \item Input: Encoded representation of the embedding.
    \item Output: Validity of the encoding.
\end{itemize}
We performed the training of the complete AAE model for 5000 epochs with a batch size of 200. For each batch the following three steps are performed:\\ \\

\textbf{-Train Discriminator:} We fed 100 training samples to the encoder and assumed that the obtained latent codes were fake (label=0). We also generated 100 samples from the desired distribution and assumed them to be real (label=1). The real latent variables for the encoder are normally distributed. After generating the latent codes, we trained the discriminator with these samples and their corresponding labels. The Adam optimization is used with a learning rate of 0.002 to fit the model. We used binary-cross-entropy for the loss function.\\ \\
\textbf{-Train Autoencoder:} The 200 samples of training embeddings are fed to the autoencoder (encoder and decoder), and the autoencoder is trained based on reconstruction error (MSE).\\ \\
\textbf{-Train Encoder:} The generator (encoder) must be trained to generate the latent codes. For this purpose, we lock the discriminator weights and trained the encoder and discriminator together so that the discriminator is fooled to classify the latent codes of the embeddings as real ones.
After training was done, a 32-dimensional feature vector was extracted from the encoder output.

Although we utilized word embedding for training these models, the extracted feature vectors provide document-level embeddings. As we illustrated above, several reshaping, sampling, and concatenation take place during the feature compaction phase. Hence, the new feature representations do not carry the word-level information anymore. Rather, they summarize the whole document.

\subsection{BERT}
\begin{figure}[!t]
\centering
\includegraphics[scale=0.4]{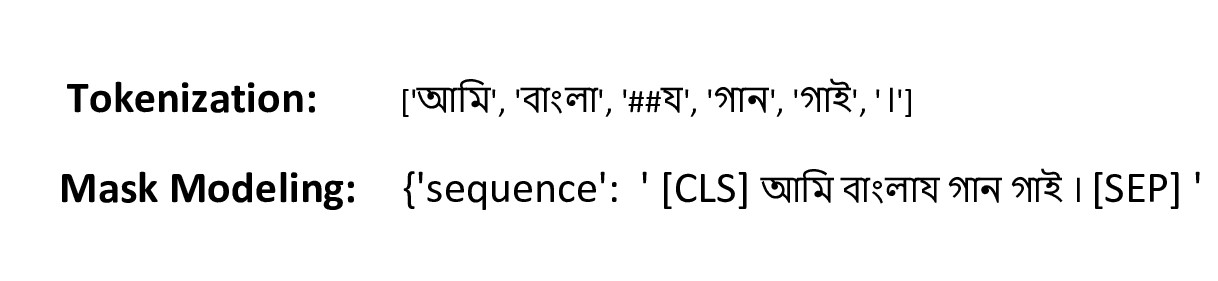} 
\caption{A sample tokenization and mask modeling by pretrained Bangla BERT}
\label{bert_bangla}
\end{figure}

Besides these three deep learning generative models that have their original application in computer vision, we also utilized a popular language model called Bidirectional Encoder Representations from Transformer (BERT). BERT represents a general language modeling that supports transfer learning and fine-tuning on specific tasks. However, in this work, we only used the feature extraction side of BERT by obtaining word embeddings \citep{reimers2019sentence} \citep{alsentzer2019publicly} from it. We utilized the BERT base model, which has 768 hidden layers with 12 transformer blocks and 12 attention heads. For tokenizing and masked language modeling, we used Bangla BERT \href{https://huggingface.co/sagorsarker/bangla-bert-base}{(https://huggingface.co/sagorsarker/bangla-bert-base)} from the Huggingface model hub. Figure \ref{bert_bangla} shows a sample tokenization and mask modeling by the pre-trained Bangla BERT model. As we mentioned earlier, the BERT base model uses 12 layers of transformer encoders. Each output per token from each of these layers can be used as a word embedding. Although the performance difference is not that much, we empirically identified that one of the best-performing choices was to sum the last four layers. We also experimented with different dimensions of embeddings up to 768, which is the maximum embedding dimension for BERT. Outcomes of these experimentations are illustrated in section \ref{res}.
\subsection{Training Classifiers}
We train three types of deep learning models as the benchmark classifiers in our work: Recurrent Neural Networks (RNNs), Convolutional Neural Networks (CNNs), and Recurrent Convolutional Networks. \\ \\
\textbf{RNNs:} RNNs have applications in text classification tasks. LSTM  \cite{hochreiter1997long} and Bidirectional LSTM \cite{schuster1997bidirectional} are the two variants of RNN that are widely used for their ‘memory’ features which can efficiently capture sequential information.	Besides the classic LSTM architecture, we also used the bi-LSTM, which puts two independent LSTM networks together in order to have both backward and forward information at every time step. We used bi-LSTM with multi-head attention, similar to the work of \cite{liu2019bidirectional}.
\\ For each of the models, we experimented by tuning the number of hidden layers from 4 to 8 and varied the number of corresponding recurrent units in each hidden layer from 32 to 256. Apart from that, We also experimented with different learning rates, batch sizes, number of epochs, dropout regularization, and number of heads in case of attention.\\
\textbf{CNNs}: Traditionally, CNNs are thought to be specialized in processing a grid of values such as image data \cite{krizhevsky2012imagenet}. The idea behind using CNNs in NLP \cite{kim2014convolutional} is to make use of their ability to extract features. In this work, the CNN model is constructed by stacking four 1D Convolution layers, interleaved with four 1D Max-pooling layers, each having a pool size of 2, followed by one 1D Global average pooling layer. For each convolution layer, we adjusted the number of filters up to 512 and the window length up to 7 based on the model’s performance.\\
\textbf{C-LSTM:} CNNs and RNNs are two mainstream architectures that adopt different ways to understand natural languages. CNNs are able to learn local responses from temporal or spatial data but cannot learn sequential correlations. In contrast, RNNs are specialized for sequential modeling but unable to extract features in a parallel way. Combining the strengths of both architectures led to the concept of recurrent convolutional networks. ~\cite{zhou2015c} proposed a simple end-to-end, unified model called C-LSTM (Convolutional LSTM) for sentence representation and text classification. They utilized CNN to extract a sequence of higher-level phrase representations and then fed this into an LSTM recurrent neural network to obtain the sentence representation. To develop the C-LSTM model, we stacked three 1D Convolution layers interleaved with the same number of 1D Max-pooling layers, followed by a flatten layer, the output of which is then fed to the first layer of a series of four LSTM layers.

We used the Adam optimizer and the binary-cross-entropy loss function for all of these deep learning models with softmax activation. During the training of the deep learning models, we applied weighted cross-entropy loss \cite{imbalanced}. Although the sample distribution in our dataset is quite uniform among the five classes, it is slightly skewed towards the \textbf{government \& politics} category. We put weights on each of the classes during the training phase so that classes with fewer samples get higher weights. We also consider macro validation recall instead of validation accuracy as the optimization metric during the training.
\begin{figure}[!t]
\centering
\includegraphics[scale=0.6]{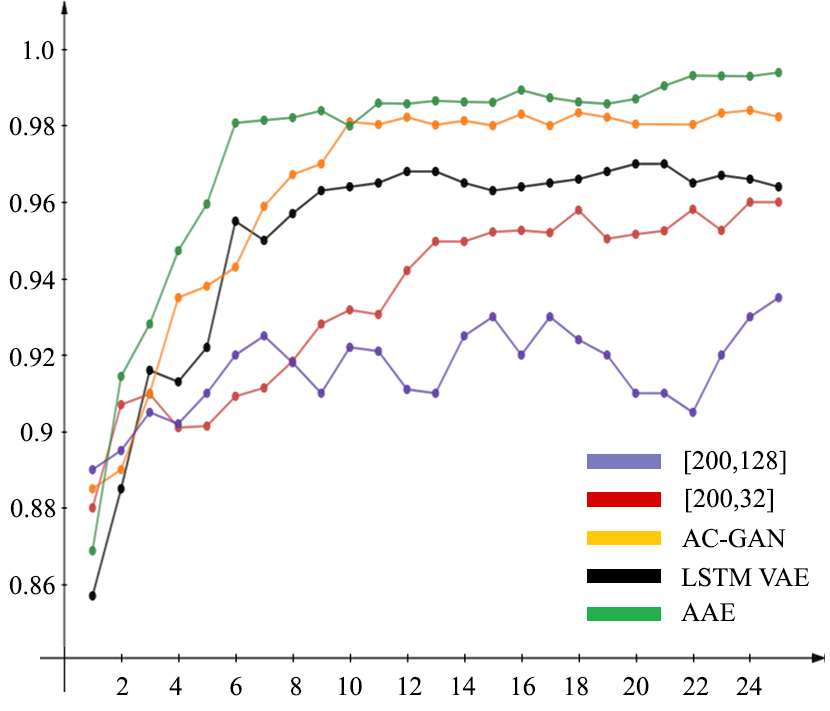} 
\caption{Validation Accuracy}
\label{class_acc}
\end{figure}

\section{Results \& Discussion}
\label{res}

To analyze the quality of the extracted feature space, our idea is to use them in some state-of-the-art NLP tasks. Hence we utilized these features in training and evaluating a couple of text classification models and later observed the classification result both for the original and the new feature representations. Since the classification models and their architectures have been kept identical in both cases, better classification scores are supposed to be derived by better feature representations.

In our case, the original vector dimension for each document is specified as $[200,128]$, and the dimension of the extracted feature space from the VAE, AC-GAN, and AAE is $[200,32]$. Therefore, we trained our classifier with the original [200,128] dimensional data vectors and the data from the three deep learning models of the reduced dimension $[200,32]$. For comparison, we also trained the classifier models with another feature space of dimension $[200,32]$, which was obtained by applying PCA \citep{pca} on the original feature space.
We use three evaluation metrics, precision, recall, and F1-score, to assess the performance of the classifiers. We use the macro average for all three metrics, i.e., the arithmetic means of all precision, recall, and F1 scores for the seven document categories. The best models for each architecture were tuned with respect to the validation set. Their performance on the test sets is shown in Table \ref{results}.

\begin{figure}[!t]
\centering
\includegraphics[width=\linewidth]{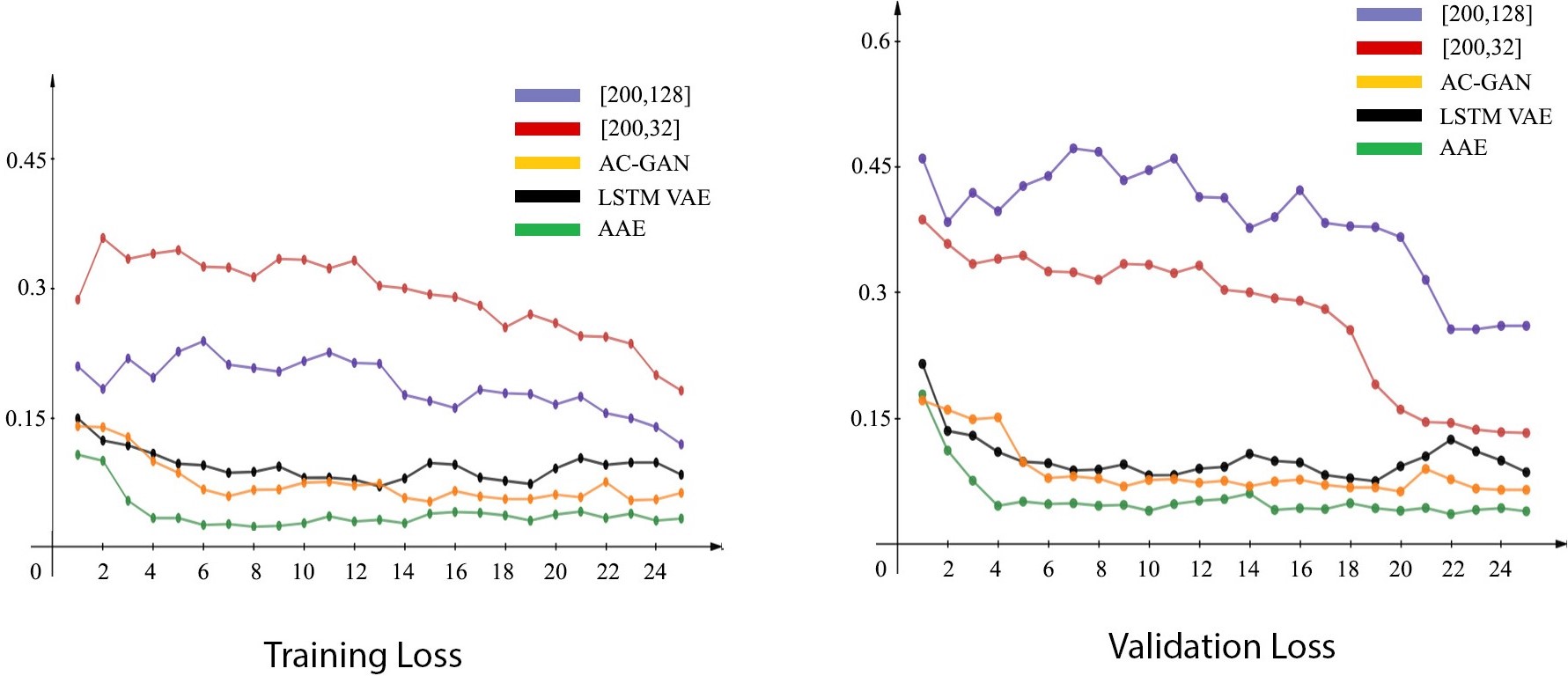} 
\caption{Training and Validation loss}
\label{class_loss}
\end{figure}

\begin{table}[!t]
\caption{Results of Text Classification models}
\begin{center}
\footnotesize{
\begin{tabular}{c|c|c|c}

     \hline
        \textbf{Feature Space} &\textbf{\%Precision} &\textbf{\%Recall} & \textbf{\%f1-score}\\ [5pt]
       \hline
       \multicolumn{4}{c}{\textbf{LSTM}}\\[3pt]
       \hline
        $[200,128]$& 85.94& 86.53& 86.23\\[1pt]
      
      PCA & 87.21& 83.28& 85.20\\[1pt]
      
      LSTM VAE& 89.12& 90.15& 89.63\\[1pt]
      
      AC-GAN& 90.31& 92.02& 91.16\\[1pt]
     
      AAE& \textbf{91.94}& \textbf{92.25} & \textbf{92.09}\\[3pt]
      \hline
       
      \multicolumn{4}{c}{\textbf{bi-LSTM}}\\[3pt]
      \hline
        $[200,128]$& 88.68& 89.91& 89.29\\[1pt]
      
      PCA & 91.55& 87.03 &89.23 \\[1pt]
      
      LSTM VAE& 93.6& 94.37& 93.98\\[1pt]
      
      AC-GAN& 93.82& \textbf{95.07}& 94.44\\[1pt]
     
      AAE& \textbf{94.44}& 94.82 & \textbf{94.63}\\[3pt]
      \hline
      
      \multicolumn{4}{c}{\textbf{bi-LSTM with Attention}}\\[3pt]
      \hline
        $[200,128]$& 89.67& 89.88& 89.77\\[1pt]
      
      PCA & 94.44& 89.07& 91.68\\[1pt]
      
      LSTM VAE& 95.33& 95.90& 95.61\\[1pt]
      
      AC-GAN& 96.47& 97.06& 96.76\\[1pt]
     
      AAE& \textbf{98.18}& \textbf{98.66} & \textbf{98.42}\\[3pt]
      \hline
      \multicolumn{4}{c}{\textbf{CNN}}\\[3pt]
        \hline
        $[200,128]$& 88.74& 90.10& 89.41 \\[1pt]
      
      PCA & 92.14& 87.80 & 89.92 \\[1pt]
      
      LSTM VAE& 94.26& 94.35& 94.30 \\[1pt]
      
      AC-GAN& 94.57& 96.13& 95.34\\[1pt]
     
      AAE& \textbf{96.73}& \textbf{97.52} & \textbf{97.12}\\[3pt]
      \hline
      
      \multicolumn{4}{c}{\textbf{C-LSTM}}\\[3pt]
      \hline
       $[200,128]$& 88.40& 89.37& 88.88\\[1pt]
      
      PCA & 90.83& 87.60 & 89.19\\[1pt]
      
      LSTM VAE& 93.74& 94.82& 94.28\\[1pt]
      
      AC-GAN& 94.02& 95.50& 94.75\\[1pt]
     
      AAE& \textbf{94.80}& \textbf{96.02} & \textbf{95.01}\\[3pt]
      \hline
\end{tabular}
}
\end{center}
 \label{results}
\end{table}

Figure \ref{class_loss} show the training and validation loss respectively for the CNN classifier model for each of the feature spaces mentioned above.  It can be seen that the validation loss is significantly higher than the training loss for the original $[200,128]$ dimensional data, indicating the prevalence of overfitting. The model starts to memorize the large feature space of the training samples and fails to predict classes for new instances accurately. Figure \ref{class_acc} depicts the validation accuracy for each of the five feature spaces. Feeding a set of embeddings of dimension [200,32], obtained by applying PCA, to the model results in somewhat better outcomes during the training session as the overfitting complication is eliminated. However, it creates a new issue, the underfitting problem. It means that the classifier does not get adequate useful features to learn from, which is necessary to distinguish the text instances based on their class. Training the model with the feature vectors of dimension [200,32], obtained with the three feature extraction models resulted in a more balanced performance on the training and validation data (Figure \ref{class_loss}), i.e., closer to the sweet spot of a good fit.         

Next, we evaluated the extracted features' usefulness by analyzing the classifier models' performance on our test data. Table \ref{results} shows the performance measures of all the document classification models for each of the five feature spaces with respect to precision, recall, and F1-scores. We can see the effect of the bad fit for the original embeddings and PCA in the test set performance as well. Each of the classifiers presents low precision and recall scores with the original embedding. While the precision scores for the PCA features are somewhat better than for the original embeddings, the recall scores drop conspicuously due to the underfitting issue, resulting in low F1 scores for all classifiers.
Looking at the results derived from the three feature extraction algorithms, it is clear that that dimensionality reduction can improve the performance of the classifier models. The LSTM VAE performs a little worse than the other two but still has an excellent set of scores, including a recall value of almost \textbf{96\%} while using the bi-LSTM model with attention, and an F1 score of more than \textbf{94\%} with CNN. 

On the other hand, transferring the AC-GAN architecture from computer vision to natural language processing did not have a bad impact at all. The bi-LSTM with attention classifier achieved more than \textbf{96\%} for precision, recall (\textbf{97.06\%}), as well as F1-score with the AC-GAN, generated feature space. 
As we mentioned earlier, the VAE model explicitly learns likelihood distribution through a loss function while the GAN learns through the so-called 'min-max two-player game.' In order to beat the generator model and improve the accuracy in classifying the real and fake sequences, the discriminator model tends to process a much useful and compact representation of the input sequences. This tendency is found somewhat less in such static loss minimization approach of the VAE models. That is why the feature space of AC-GAN outperformed that of LSTM VAE.

Nevertheless, the adversarial autoencoder model outperformed both of them. It has a precision of more than \textbf{98\%} and recall of nearly \textbf{99\%} (\textbf{98.78\%}), resulting in the best F1-score of \textbf{98.42\%} for the bi-LSTM model with attention. In fact, its feature space led to the best results for all other classification models as well, including an F1-score of more than \textbf{97\%} with CNN, and a recall value of more than \textbf{96\%} with the C-LSTM model. The adversarial autoencoder model effectively merges the good properties of the VAE and the GAN models, making it a better performing architecture than the other two. The adversarial loss optimization of AAE through the discriminator, unlike the KL divergence loss of LSTM VAE enables it to produce better generative performance. At the same time, generating the latent codes instead of fake sequences from the generator(encoder) makes the model more competent than GAN in processing the internal feature representations.

We kept the architectures of the classifiers unchanged for each of these three feature extraction models so that we can precisely evaluate the impact of their feature space in the classification task. Among the RNN based classifiers, LSTM derived much lower scores than the other biLSTM models. It is not surprising because bi-LSTM models preserve the information from both forward and backward directions, which helps in improving the prediction in such sequence classification task. However, adding multi-head attention layers to the biLSTM model derived the best results in this work, since this mechanism cognitively enhances the important segments of the feature space and fades out the rest. Apart from that, CNN model delivered a better prediction accuracy than biLSTM. With the AAE feature space, it derived more than 97\% f1, which indicates that its local filtering over the abstract feature space helps significantly during the prediction. Finally, the blending of CNN and LSTM model, i.e., C-LSTM outperformed both LSTM and biLSTM models but could not add any improvement to the results of CNN.

\begin{table}[!t]
\centering
\small
\begin{tabular}{|c|c|c|c|c|c|c|c|c|c|c|}
\hline  
Embedding&  Model&  P&  R&  F1& P& R& F1& P& R& F1       \\ 
  Dimension&&&&&&&&&&\\
\hline  
                  &  & \multicolumn{3}{c|}{Sum All 12 Layers} & \multicolumn{3}{c|}{Sum Last 4 Layers} & \multicolumn{3}{c|}{Only Last Layer} \\ \hline
  &  BiLSTM-attention & 97.72 & 97.25 & 97.48 & \textbf{99.25} & \textbf{98.87} & \textbf{99.06} &98.31 & 97.92 & 98.11\\ \cline{2-11} 
               
   768&  CNN & 95.83 & 96.17 & 96.00 & 97.64 & 97.22 & 97.43 &95.53 & 95.85 & 95.69\\ \cline{2-11} 
                 
   &  C-LSTM & 93.47& 94.28& 93.87 &  95.32 & 94.95 & 95.13 & 94.72 & 94.44 & 94.58 \\ \hline
                  
  &  BiLSTM-attention & 94.37 & 94.25 & 94.29 & 95.25 & 94.87 & 95.10 &94.81 & 94.52 & 94.69\\ \cline{2-11} 
                
   450 & CNN& 95.33 & 94.98 & 95.17 & \textbf{96.10} & \textbf{95.88} & \textbf{95.99} &95.77 & 95.54 & 95.64 \\ \cline{2-11} 
                 
  &C-LSTM & 91.74 & 91.25 & 91.49 & 93.05 & 92.77 & 92.90 &92.31 & 92.02 & 92.19\\ \hline            
                
  &  BiLSTM-attention & 91.67 & 91.36 & 91.5 & 92.33 & 92.07 & 92.21 &92.10 & 91.82 & 91.97\\ \cline{2-11} 

  200 &  CNN& 92.24 & 92.00 & 92.11 & \textbf{93.19} & \textbf{92.97} & \textbf{93.09} &92.68 & 92.37 & 92.53 \\ \cline{2-11} 
               
  & C-LSTM& 88.97 & 88.66 & 88.80 & 90.63 & 90.35 & 90.50 &91.10 & 90.89 & 90.97       \\ \hline
                  
\end{tabular}
\caption{experimental results with BERT embeddings}
\label{bert_result}
\end{table}

\begin{table}[!t]
\centering
\begin{tabular}{|l|l|l|l|l|}
\hline
 df   & sum\_sq & mean\_sq & F\_value & p\_value \\ \hline
 5.00  & 171.32 & 34.26   & 19.87   & 0.00002  \\ \hline
 12.00 & 20.69   & 1.73    & NaN      & NaN      \\ \hline
\end{tabular}
\caption{ANOVA Table}
\label{anova}
\end{table}

\begin{table}[!t]
\centering
\footnotesize{
\begin{tabular}{c|c|c|c|c}

     \hline
       \multirow{2}{*}{Dataset}
       &\multirow{2}{*}{Models} 
       &\multirow{2}{*}{\%Precision} &\multirow{2}{*}{\%Recall} &\multirow{2}{*}{\%F1-score}\\
       & & & &\\
      \hline
      ProthomAlo + BDNews24 & TF-IDF + DenseNN & 86.00 & 85.00 & 84.00\\
      \cite{prothombdnews}& & & &\\
      & & & &\\
      \hline
      \multirow{2}{*}{BARD} & CNN & --- & --- & 91.79\\
      & LSTM& --- & --- &90.76\\
      \cite{bardprothom} & & & &\\
      \multirow{2}{*}{ProthomAlo} & CNN & --- & ---& 91.08\\
      & LSTM& --- & ---& 92.57\\
      \hline
      
      \multirow{4}{*}{BDNews24} & Ridge Classifier & 93.61 & 93.45 & 93.44\\
      & Passive-Aggressive & 93.21 & 93.23 & 93.19\\
      & SVM & 93.80 & 93.83 & 93.78\\
      \cite{bdnews} & Logistic-Regression & 93.11 & 93.10 & 93.05\\
      & SGD Classifier & 93.86 & 93.88 & 93.85\\
      \hline
      
      AnandaBazar + Bartaman & Cosine similarity & 95.80 & 95.80 & 95.80\\
      + Ebela tabloid & Euclidian distance & 95.20 & 95.20 & 95.20\\
      \cite{ananda} & & & &\\
      & & & &\\
      \hline
      
      BD Corpus & SVM & --- & ---& 89.14\\
      \cite{bdcorpus} & NB classifier & --- & ---& 85.22\\
      & & & &\\
      \hline
      
      BARD & Word2Vec+ Logistic regression & 95.00 & 95.00& 95.00\\
      \cite{bard} & & & &\\
      & Word2Vec+ Neural Net & 96.00 & 96.00& 96.00\\
      \hline
      
      Our Corpus & AAE + BiLSTM with Attention & \textbf{98.18} & \textbf{98.66} & \textbf{98.42}\\
      & & & &\\
      \hline
      
\end{tabular}
}
\caption{Existing works along with our proposed model}
 \label{comparison}
\end{table}

We also evaluated the performance of BERT embedding in the text classification task. Being a language model, BERT is supposed to extract a better feature space from text compared to the other generative models we used in this work. In Table \ref{bert_result} we demonstrate the experimental results derived by training the text classifiers with BERT embeddings. The embeddings were obtained by summing the outputs of different layers. Here we mentioned about three approaches, i.e., summing all the 12 layers, summing the last 4 layers, and considering only the last layer. We also experimented by reducing the original embedding dimension (768) of BERT to 450 and 200 respectively. The results of LSTM and biLSTM models are excluded from Table \ref{bert_result} , and only the scores of the three best-performing models are mentioned. We can see that summing the outputs from all 12 layers did not perform as well as the other approaches. Since the internal feature representations(embeddings) at the initial layers are less useful and more abstract in nature, their inclusion in the summation process basically degrades the resultant embeddings. In another extreme, if we take only the last layer's output as the embeddings, it is not the best, although it shows a better performance than the previous approach. We found the best set of scores by summing the outputs of the last 4 transformer layers of the BERT base model. On the other hand, embeddings with reduced dimensions seem to work worse than the original one. The reason is that the BERT model does not facilitate any custom dimensionality reduction by default. To get the feature space in the lower dimension (i.e., 450 or 200), we naively truncated the original embeddings into the smaller ones. It basically caused some information loss which had a clear impact on the document classification task. However, the best outcome of \textbf{99.06\%} f1-score is produced by the biLSTM with attention model while summing the last four layers with the original embedding dimension. Nevertheless, with the lower embedding dimensions (e.g., 450 and 200), CNN outperformed the other classifiers. Since the original embedding was truncated, the RNN based models could not preserve much useful information while parsing the sequences. In such a scenario, the local filtering scheme of CNN turned out to be more effective. Although the BERT embeddings have generated the best feature space as expected, our proposed generative models, specially AAE and AC-GAN, do not lag much behind. Rather, the implication of their feature spaces in the document classification task is very much comparable, which we can see from Table \ref{results} and \ref{bert_result}.  

From the overall results mentioned above, we found that several models in our experiments performed closely. Hence we undertook the ANOVA test \citep{st1989analysis}, a statistical significance testing strategy that helps to find out whether the differences between groups of data (results in our case) are statistically significant. We conducted the one-way ANOVA test among the f1-scores derived by the original embedding, PCA, VAE, AC-GAN, AAE, and BERT embeddings. For all of them, we only considered the results of the top three performing models, i.e., BiLSTM-attention, CNN, and C-LSTM. So, the test was done on three f1 scores for all the feature extraction methods mentioned above. Table \ref{anova} shows the outcome of the Anova test. Since The p-value obtained from ANOVA analysis is statistically significant (< 0.05), therefore, we conclude that the results of our proposed models are effective. 

From the aforementioned analysis, we know that score differences are statistically significant, but ANOVA does not tell us which results are significantly different from each other. To know the pairs of significantly different scores, we also performed multiple pairwise comparison (post hoc comparison) analysis using Tukey’s honestly significant difference (HSD) test \citep{abdi2010tukey}. Results from Tukey’s HSD test is shown in \ref{tukey}.

In Table \ref{comparison} we demonstrate a series of existing works along with their classification models and results. Most of the existing works have utilized TF-IDF and Word2Vec as text features. Besides, they trained several state of the art machine learning classifiers for the task. However, almost all of these works have been done on some small-sized datasets, where a couple of them have a large enough text collection with a very few number of domains and sources. On the other hand, we developed a comprehensively large dataset distributed among seven categories for this task.

\section{Conclusions}
\label{conc}
We presented a comprehensive human-annotated dataset of Bangla articles and implemented three deep learning generative models to extract textual features. We trained these models with our curated dataset and utilized the feature space obtained with the three models for document classification. We analyzed the usefulness of these text features by evaluating the performance of the classifiers and found that the adversarial autoencoder model produced the best feature space. As a future extension of this work, we intend to use the deep learning generative models for Bangla text completion and summarization tasks. Analyzing the performance of recent generative models such as Nouveau VAE (NVAE) \cite{2020arXiv200703898V}, and Wasserstein Auto-Encoders \cite{bahuleyan2018stochastic} for feature extraction in different NLP tasks could be another potential avenue for future research. We will also continue to expand our dataset and extend the number of text categories. Besides, this dataset is made available for public use (\href{https://cutt.ly/SYTV6Pv}{https://cutt.ly/SYTV6Pv}). We hope it will help the research community to carry out further research on Bengali language processing. 

\section{Acknowledgements}
We are thankful to our data annotator team and Samsung Applied Machine Learning Laboratory at BUET, where the experiments were conducted.
\bibliography{main}
\hfill

\appendix

\section{CORPUS}
\label{corpusA}
Description of the original dataset with related metadata can be found in the following link :
\href{https://drive.google.com/drive/folders/11NMs94gF4sJxFstOYylLKLKCz0vCqZ1c?usp=sharing}{\textcolor{blue}{csv file link}}

\section{TEXT SOURCES}
\label{sourceA}

\begin{longtable}{c|c|c}

     \hline
       \multirow{3}{*}{} \textbf{Source} & \textbf{Alexa Rank} & \textbf{\#Text Document}\\
       &&\\
      \hline
      \multicolumn{3}{c}{\textbf{Govt. \& Politics}}\\
      \hline
      www.prothomalo.com & 548 & 21267 \\
      www.anandabazar.com & 3352 & 12466 \\
      www.kalerkantho.com & 1632 & 11112  \\
      www.jagonews24.com & 1647 & 6032 \\
      www.samakal.com & 6990 & 1189 \\
      www.jugantor.com & 1166 & 3739 \\
      bangla.bdnews24.com & 1778 & 1566 \\
      www.ntvbd.com & 9579 & 3155 \\
      www.banglatribune.com & 2898 & 2448 \\
      \hline
      \multicolumn{3}{c}{\textbf{Science \& Technology}}\\
      \hline
      www.prothomalo.com & 548 & 3457 \\
      www.anandabazar.com & 3352 & 3236 \\
      www.banglatech24.com & 166875 & 4311  \\
      \hline
      
      \multicolumn{3}{c}{\textbf{Economics}}\\
      \hline
      www.prothomalo.com & 548 & 21267 \\
      www.ittefaq.com.bd & 3841 & 2448 \\
      www.kalerkantho.com & 1632 & 11112  \\
      www.jagonews24.com & 1647 & 6032 \\
      www.dhakatimes24.com & 21007 & 3739 \\
      bangla.bdnews24.com & 1778 & 1566 \\
      www.samakal.com & 6990 & 1189 \\
      \hline
      
      \multicolumn{3}{c}{\textbf{Health \& Lifestyle}}\\
      \hline
      www.prothomalo.com & 548 & 7067 \\
      www.anandabazar.com & 3352 & 1163 \\
      www.kalerkantho.com & 1632 & 2964  \\
      www.ntvbd.com & 9579 & 1241 \\
      www.priyo.com & 35131 & 1032 \\
      \hline
      
      \multicolumn{3}{c}{\textbf{Entertainment}}\\
      \hline
      www.prothomalo.com & 548 & 16331 \\
      www.ittefaq.com.bd & 3841 & 6818 \\
      www.kalerkantho.com & 1632 & 3094  \\
      www.jagonews24.com & 1647 & 2723 \\
      www.samakal.com & 6990 & 1560 \\
      www.jugantor.com & 1166 & 1856 \\
      bangla.bdnews24.com & 1778 & 2320 \\
      \hline
      
      \multicolumn{3}{c}{\textbf{Arts \& Literature}}\\
      \hline
      www.prothomalo.com & 548 & 1106 \\
      www.anandabazar.com & 3352 & 728 \\
      www.kaliokalam.com & 851158 & 7350 \\
      www.somewhereinblog.net & 65259 & 6030 \\
      \hline
      
      \multicolumn{3}{c}{\textbf{Sports}}\\
      \hline
      www.prothomalo.com & 548 & 18450 \\
      www.kalerkantho.com & 1632 & 7433  \\
      www.jagonews24.com & 1647 & 10234 \\
      www.samakal.com & 6990 & 8830 \\
      www.jugantor.com & 1166 & 4217 \\
      bangla.bdnews24.com & 1778 & 2775 \\
      www.banglatribune.com & 2898 & 1962 \\
      \hline
\caption{News Sources}     
\end{longtable}

\section{RESULTS FROM TUKEY’S HSD TEST}
\label{tukey}
\begin{longtable}{l|l|l|l|l|l|l}

\hline
\textbf{Group 1} & \textbf{Group 2} & \textbf{Diff}               & \textbf{Lower}               & \textbf{Upper}              & \textbf{q-value}            & \textbf{p-value}               \\ \hline
orig    & pca     & 0.909 & -2.692 & 4.512 & 1.200 & 0.900                   \\ \hline
orig    & aae     & 7.496   & 3.894  & 11.098 & 9.887  & 0.001                 \\ \hline
orig    & vae     & 5.376  & 1.774  & 8.978  & 7.091  & 0.003  \\ \hline
orig    & gan     & 6.263  & 2.661  & 9.865  & 8.260  & 0.001                 \\ \hline
orig    & bert    & 7.853  & 4.251    & 11.455  & 10.357 & 0.001                 \\ \hline
pca     & aae     & 6.586  & 2.984   & 10.188 & 8.687  & 0.001                 \\ \hline
pca     & vae     & 4.466  & 0.864  & 8.068  & 5.891  & 0.012   \\ \hline
pca     & gan     & 5.353  & 1.751  & 8.955  & 7.060 & 0.003 \\ \hline
pca     & bert    & 6.943  & 3.341   & 10.545 & 9.157  & 0.001                 \\ \hline
aae     & vae     & 2.120 & 1.482 & 5.722 & 2.796 & 0.040    \\ \hline
aae     & gan     & 1.233 & 2.368 & 4.835  & 1.626 & 0.084    \\ \hline
aae     & bert    & 0.356 & 3.245   & 3.958   & 0.470 & 0.090                   \\ \hline
vae     & gan     & 0.886 & 2.715 & 4.488  & 1.169 & 0.090                   \\ \hline
vae     & bert    & 2.476 & 1.125 & 6.078  & 3.266 & 0.026   \\ \hline
gan     & bert    & 1.589 & 2.012 & 5.192  & 2.097   & 0.066      \\ \hline
\caption{Results from Tukey’s HSD}
\end{longtable}

Results from Tukey’s HSD suggests that almost all the pairwise comparisons rejects null hypothesis (p-value $< 0.05 $) and indicates statistical significant differences.
\end{document}